\documentstyle[a4,11pt]{article}

\newcommand{\eqalign}[1]{
\null \,\vcenter {\openup \jot \ialign {\strut \hfil $\displaystyle {
##}$&$\displaystyle {{}##}$\hfil \crcr #1\crcr }}\,}

\newcommand{\bd}{\begin{document}}
\newcommand{\ed}{\end{document}}
\newcommand{\bc}{\begin{center}}
\newcommand{\ec}{\end{center}}
\newcommand{\bfr}{\begin{flushright}}
\newcommand{\efr}{\end{flushright}}
\newcommand{\bfl}{\begin{flushleft}}
\newcommand{\efl}{\end{flushleft}}
\newcommand{\be}{\begin{equation}}
\newcommand{\ee}{\end{equation}}

\newcommand{\IR}{{\bf R}}
\newcommand{\IZ}{{\bf Z}}
\newcommand{\szm}{$\sigma$ mo\-del}
\newcommand{\pa}{\partial}
\newcommand{\bs}{\backslash}

\title{A Two-loop Test of Buscher's T-duality I}
\author{Zal\'an~Horv\'ath$^1$\thanks{e-mail address:
zalan@kvark.elte.hu},\,\,
Robert L.~Karp$^{1,2}$\thanks{e-mail address: karp@physics.uc.edu},\,\,
L\'aszl\'o~Palla$^1$\thanks{e-mail address: palla@ludens.elte.hu}\\
\\
\it \small ${}^1$  Institute of Theoretical Physics, E\"otv\"os
University, Budapest, Hungary\\
\it \small
${}^2$ Physics Department, University of Cincinnati, Cincinnati, OH 45221}
\date{}

\begin{document}

\maketitle

\vspace{1in}

\begin{abstract}
We study the two loop quantum equivalence of sigma models related by
Buscher's T-duality transformation. The computation of the two loop
perturbative free energy density is performed in the case of a certain
deformation of the $SU(2)$ principal sigma model, and its T-dual, using
dimensional regularization and the geometric sigma model perturbation
theory. We obtain agreement between the free energy density expressions of
the two models. 
\end{abstract}

\vspace{11 mm}
\begin{center}
key words: sigma models, duality, quantum corrections
\end{center}

\newpage

\section{Introduction}

Among the wealth of different dualities relating the perturbative string
theories and M-theory, the most 'ancient' one is T-duality. At the same
time T-duality was the starting point in the discovery of D-branes
\cite{Polchinski}. In string theory T-duality can be proven in arbitrary
order of the string perturbation theory \cite{poratti,polchinski}, as long
as the vacuum preserves conformal invariance \cite{Rocek}. Generalizing
the $R\longrightarrow{1\over R}$ symmetry of the toroidally compactified
strings, Buscher \cite{busch} gave a set of equations that describe the
transformation of the Neveu-Schwarz background fields of the sigma model
action, and works whenever there is an isometry. Denoting by $g_{\mu\nu}$
and $b_{\mu\nu}$ the background metric and the antisymmetric tensor field,
the explicit form of the Buscher transformation is (since we work on a
flat world sheet, we neglect the dilaton): 
\be\eqalign{ 
&{\tilde g}_{00}={1\over g_{00}},\qquad {\tilde g}_{0\alpha}={b_{0\alpha}
\over g_{00}}, \qquad {\tilde b}_{0\alpha}={g_{0\alpha} \over g_{00}}, \cr
&{\tilde g}_{\alpha\beta}=g_{\alpha\beta} - {g_{0\alpha}g_{0\beta} -
b_{0\alpha} b_{0\beta}\over g_{00}},\quad {\tilde
b}_{\alpha\beta}=b_{\alpha\beta}-{g_{0\alpha}b_{0\beta}
-g_{0\beta}b_{0\alpha}\over g_{00}}, \cr} 
\ee 
where ${\tilde g}_{\mu\nu}$
and ${\tilde b}_{\mu\nu}$ denote the new background fields.

To derive this transformation Buscher used functional integral arguments,
that in the meantime have become widely known, and applied in many context
(see e.g. \cite{witten}). The idea is to gauge the aforementioned isometry
of the sigma model action and impose a constraint using a Lagrange
multiplier. Integrating the multiplier one recovers the original theory to
start with, while after a gauge fixing and integrating over one of the
original fields gives the dual theory.

It was also shown \cite{ealv}, that, at the classical level, the duality
transformation rule can be recovered in an elegant way by performing a
canonical transformation \cite{lozano}. This clearly shows that the models
connected by the Buscher transformations are equivalent classically. At
the quantum level the only loophole of the path integral argument
mentioned above is that it neglects the effect of the regularization and
renormalization. Though it was shown in \cite{Rocek} that for conformal
invariant models one has full quantum equivalence, in other words the path
integral argument holds, this is not the case in the non-conformal
setting. It was shown in \cite{bfhp} for a deformed $SU(2)$ principal
sigma model, and further clarified in \cite{en} for several different
deformed $SU(3)$ principal sigma models, that Buscher's formula -- as
applied to renormalized quantities -- has to be modified to give two loop
quantum equivalence.

Later it was shown that the one loop beta functions of the original and
dual models always agree \cite{haagensen1}. Work has been done to
establish the corrections to Buscher's formulae
\cite{bfhp1,balog1,haagensen2}. Advances were made from a different point
of view. It is widely known that the low energy degrees of freedom of the
sigma model can be described using an effective action, that contains
gravity in target space. This fact constrains the possible leading terms
of the low energy effective action to a computable form, that is known. It
was shown \cite{kaloper1} that the low energy effective action consistent
with the two loop sigma model beta function equations is not invariant
under Buscher's transformation. The leading part of the above action, the
one loop part, nevertheless is invariant, in accordance with the one loop
findings of \cite{haagensen1}. Using the non-invariance of the two loop
action it was found how to modify Buscher's transformation, with order
$\alpha'$ terms, such that the two loop action would remain invariant
under the modified transformations.  What is not clear after all is how to
pull back the modification of the Buscher's formulae found in the low
energy action to the sigma model. Related work in the supersymmetric case,
concentrating on the absence of the above mentioned corrections, was done
in \cite{alvarez,penati}.

Any satisfactory criterion of the quantum equivalence among dually related
sigma models should be based on the comparison of physical quantities as
opposed to just considering beta functions. If there are global symmetries
in the model then their associated conserved quantities (Noether currents)
may be considered physical.  The definition of physical quantities,
however, is not very clear in diffeomorphic invariant sigma models without
a sufficient number of isometries. To circumvent this problem the study of
Weyl anomaly coefficients was suggested in \cite{balog1}. In the present
paper -- as an alternative -- we study a thermodynamic quantity namely the
free energy density in the presence of a chemical potential in the dually
related sigma models. This quantity surely qualifies as physical, thus its
equality in the two models gives a non trivial check on their quantum
equivalence. Furthermore the free energy density can be computed
perturbatively -- at least in asymptotically free models, thus one can
compare the two free energy densities using the first few orders of
perturbation theory. The aim of this paper is to carry out this comparison
in the two loop order, where the first really \lq quantum' effects appear,
thus improving the almost \lq classical' one loop case studied earlier in
\cite{KP1}, \cite{pall2}.
   
The paper is organized as follows: in Section 2 we give a brief review of
the pertinent facts that we need from the renormalization of the $SU(2)$
principal {\szm } and its dual and develop a Lagrangian with more
parameters, that can accommodate both Lagrangians as special cases. In
Section 3 we investigate the conserved charges that can be coupled to the
models, calculate the Hamiltonian, find the ground states in the presence
of the external field, and make the Lagrangians suitable for a
perturbative computation.  Section 4 deals with the definition of the
perturbative free energy density, and its computation, using dimensional
regularization. First we define the perturbative free energy density, and
set up a scheme to compute it systematically to any order.  The rest of
Section 4 deals with the actual computation of the bare free energy up to
two loops.  In Section 5 we deal with the issue of renormalization and
obtain the one and two loop renormalized free energies of the original and
dual models, compare them, and improve them using the renormalization
group. We close Section 5 with the analysis of the composite operator
renormalization of the relevant operators. We make our conclusions in
Section 6.

\section{Lagrangians and T-duality}

\subsection{The deformed $SU(2)$ principal \szm}

This section has a twofold role. Primarily it is intended to give an
overview of the results that we need in the rest of the paper and at the
same time fix the notations. Secondarily it extends some of the earlier
results in a way suitable for our applications.

In \cite{bfhp} the following one parameter deformation of the $SU(2)$
principal $\sigma$-model was considered :
\be
{\cal L}_O=-{1\over2\lambda}\bigl(\sum\limits_{a=1}^3J_\mu^aJ^{\mu a}+
 gJ_\mu^3J^{\mu 3}\bigr)\,, 
\label{lagrsig}
\ee 
where $J_\mu=G^{-1}\pa_\mu G=J_\mu^a\tau^a$, and $\tau^a=\sigma^a/2$ with
$\sigma^a$ being the standard Pauli matrices.  Thus $G$ is an element of
$SU(2)$ and $g$ is the parameter of the deformation.  From the Lagrangian
(\ref{lagrsig})  it is clear that the global $SU(2)_L\times SU(2)_R$
symmetry of the undeformed principal $\sigma$-model is broken to
$SU(2)_L\times U(1)_R$ by the $J_\mu^3J^{\mu 3}$ term.  Setting $g=0$
corresponds to the principal $\sigma$-model, while for $g=-1$ the $O(3)$
$\sigma$-model is obtained as it can be seen from eq.~(\ref{lagrsig2})  
below.

The authors of \cite{bfhp} investigated the renormalization of $\lambda$
and $g$ in the two-loop order of perturbation theory, treating $\lambda$ 
as the coupling constant and $g$ as a parameter.  Using the Euler angles
($\phi$,$\theta$,$\psi$) to parameterize the elements of $SU(2)$, $G$ is
written as
\be\label{euler}
G=e^{i\phi\tau^3}e^{i\theta\tau^1} e^{i\psi\tau^3}\,.
\ee
Using this converts the Lagrangian of the deformed \szm , which for the
time being we shall call {\em 'the original model'}, into the following
form:
\be\label{lagrsig2}
\eqalign{
{\cal L}_O=&{1\over2\lambda}\bigl\{ 
(\pa_\mu\theta)^2+(\pa_\mu\phi)^2\bigl( 1+g \cos^2\theta\bigr)+\cr
&+(1+g)(\pa_\mu\psi)^2+2(1+g)\pa_\mu\phi\pa^\mu\psi\cos\theta\bigr\}\,.
}
\ee

Using the Killing vectors of the $SU(2)_L \times U(1)_R$ symmetry and
exploiting the manifest target space covariance of the background field
method it was proved in \cite{bfhp} that the model is renormalizable in
the ordinary sense: there is no wave function renormalization for
$\theta$, $\phi$ and $\psi$, while the coupling constant and the parameter
got renormalized according to:
\be\label{couplren1}
\lambda_0=\mu^{-\epsilon} Z_\lambda(\lambda ,g)\lambda ,\qquad
g_0=Z_g(\lambda ,g)g.
\ee
Both in the one and in the two loop orders
the residues of the single poles in $Z_\lambda(\lambda ,g)=
1-y_\lambda(\lambda ,g)/\epsilon +\ldots$ and $Z_g(\lambda ,g)=
1-y_g(\lambda ,g)/\epsilon +\ldots$ were determined:
\be\label{bf1}
\eqalign{
y_\lambda=&-{\lambda\over4\pi}\bigl(
1-g+{\lambda\over16\pi}(1-2g+5g^2)\bigr)\,,\cr
y_g=&{\lambda\over2\pi}(1+g)\bigl( 1+{\lambda\over8\pi}(1-g)\bigr)\,.
}\ee
Note the sign difference between our formulas (\ref{couplren1}) and
(\ref{bf1}), and the corresponding ones in \cite{bfhp}.
It is consequence of the fact that in our notation notation $n=2+\epsilon$
rather than $n=2-\epsilon$ as used in \cite{bfhp}.

The standard definition of the $\beta$ functions: 
$\beta_\alpha=\mu{d\alpha\over d\mu}$,
$\beta_\gamma=\mu{d\gamma\over d\mu}$,
lead to the following two-loop $\beta$ functions (eq. (20) in
\cite{bfhp}):
\be\label{betasig1}
\eqalign{
\beta
_\lambda=&-{\lambda^2 \over4\pi}\bigl(
1-g+{\lambda\over8\pi}(1-2g+5g^2)\bigr)\,
,\cr
\beta_g=&{\lambda\over2\pi}g(1+g)\bigl( 1+{\lambda\over4\pi}(1-g)\bigr)\,.
}\ee

It is easy to see, that the $g=0$ resp.\ the $g=-1$ lines are fixed lines
under the renormalization group flow, and $\beta_\lambda$ reduces to the
$\beta$ function of the principal $\sigma$-model, resp.\ of the $O(3)$
$\sigma$-model on them.  In the ($\lambda\ge0, g<0$) quarter of the
($\lambda,g$) plane the renormalization group trajectories run into
$\lambda=0, g=-1$; while for $g>0$ they run to infinity. This implies that
the $g=0$ fixed line corresponding to the principal $\sigma$-model is
`unstable' under the deformation.

The Lagrangian of the deformed $\sigma$-model, eq.~(\ref{lagrsig2}),
exhibits two obvious Abelian isometries that can be used to construct two
different (Abelian) duals: namely the translations in the $\phi$ and
$\psi$ fields. We call the models obtained this way the `$\phi$ dual'
respectively the `$\psi$ dual' of the deformed $\sigma$ model
(\ref{lagrsig2}). 

In \cite{bfhp} it was found that for the `$\psi$ dual' model, as
summarized below, the renormalization of the coupling and the parameter
are equivalent to that of the original model. Therefore, in the present
context we deal with the original, deformed $SU(2)$ principal
$\sigma$-model and it's `$\psi$ dual', which we shall simply call {\em
'the dual model'}.

\subsection{The `$\psi$ dual' model}

For the Lagrangian of the '$\psi$ -dual' model, using Buscher's formulae
\cite{busch,bfhp}, one has an expression analogous to
eq.~(\ref{lagrsig2}):
\be\label{psilag}
\eqalign{
{\cal L}_D
=&{1\over2\tilde\lambda}\Bigl( (\pa_\mu\theta)^2+(\pa_\mu\phi)^2
\sin^2\theta+(\pa_\mu \chi)^2
+2a\cos\theta\epsilon^{\mu\nu}
\pa_\mu \chi\pa_\nu\phi\Bigr)\,.
}\ee
Here $\chi$ denotes the (appropriately scaled) variable dual to $\psi$, and
($\tilde\lambda,\tilde g$) stands for the couplings of the dual model.  
One can  show that 
${\cal L}_D$ exhibits the expected $SU(2)\times U(1)$ symmetry
for all values of the parameter $a$.

The couplings of the original, (\ref{lagrsig2}), and of the dual models,
(\ref{psilag}), are related (at the classical level) as a direct
consequence of T-duality as
\be\label{couplrel}
\tilde\lambda =\lambda\,,\quad a=\sqrt{1+\tilde g}\,,
\quad \tilde g=g\,.
\ee
These relations can be maintained at the two-loop level if one performs
the renormalization and, in addition, in both theories the couplings
$(g\,,\tilde g)$ are expressed in terms of the corresponding
renormalization group invariant quantities, which at the end are set equal
to each other \cite{bfhp}. In addition one also has to take into account
the freedom in the choice of the renormalization group invariant, due to
the scheme dependence of the two-loop $\beta$
function.

The two-loop renormalization invariants, that characterize the flows under
the corresponding sets of $\beta$ functions, for the original and dual
models can be easily computed:
\be\label{reninv}
\eqalign{
&M_O=-\frac{g}{(1+g)^2}\lambda^2-{1\over4\pi}\frac{g}{1+g}\lambda^3+ 
{\tt o}(\lambda^4)\,,\cr
&M_D=-\frac{a^2-1}{a^4}\tilde\lambda^2-{1\over4\pi}\frac{1}{a^2}\tilde 
\lambda^3+{\tt o}(\tilde\lambda^4)\,.
}
\ee

The next step is to express $g$ (resp. $a$) in terms of the
renormalization group invariant. This can be done by inverting
eq.~(\ref{reninv})  perturbatively. In both cases at leading order one has
to solve quadratic equations and the next to leading order terms correct
the results including the two-loop effects. Assuming that neither $M_O$
nor $M_D$ vanishes, one obtains (the leading terms were obtained in
\cite{bfhp}, but we shall need the whole expression):
\be\label{reninv1}
\eqalign{
&g(\lambda,M_O)=-1\pm{1\over\sqrt{M_O}}\lambda-{1\over2M_O}\lambda^2+ 
{\tt o}(\lambda^3)\,,
\cr
&a^2(\tilde{\lambda},M_D)=\pm{1\over\sqrt{M_D}}
\tilde{\lambda} -{1\over{2M_D}}\tilde{\lambda}^2+
{\tt o}(\tilde{\lambda}^3)\,.
}
\ee
The sign ambiguity can be removed by studying the renormalization group
flows, as it was briefly mentioned in connection with
eq.~(\ref{betasig1}). In the original model, it turns out that the
interesting region is the vicinity of $g=-1$ with $g\geq -1$. In the dual
model, since $a=\sqrt{1+g}$, choosing $g\geq -1$ one is uniquely led to
the solution with the plus sign. Thus in both cases one has to consider
the solution with the plus sign. Moreover if one sets $M_O$ and $M_D$
equal, $M_O=M_D=M$, then the classical $a=\sqrt{1+g}$ relation can be
maintained. We note (though we didn't display it in eq. (\ref{reninv1}))
that the two expressions differ already at the order of $\lambda^3$, as
expected, since $M$ is renormalization group invariant only up to two loop
order.

Using $g(\lambda,M_O)$ ($a^2(\tilde{\lambda},M_D)$) in the two loop beta
functions of the coupling constants of the original and dual models yields
a universal expression \cite{bfhp}:
\be\label{b1}
\beta_\lambda=-\frac{\lambda^2}{4\pi}\left(2+\lambda({1\over \pi} -{1\over
\sqrt{M}})\right).
\ee
As far as the coupling constant renormalization is concerned, this
universal beta function shows that the two models are equivalent, both are
asymptotically free, and the actual value of $M$ effects only the two
loop coefficient.

According to eq. (\ref{reninv1}) one can express $a$ in terms of the
common renormalization group (RG) invariant. For the sake of simplicity it
is useful
to introduce the following notations:
\be\label{fundpar}
A=\sqrt{\lambda}\,,\quad
\alpha_1= \frac{1}{\sqrt[4]{M}}\,,\quad
\alpha_2= \frac{1}{4\sqrt[4]{M^3}}\,.
\ee
It will turn out that $A=\sqrt{\lambda}$ is the proper coupling of the two
models.  In terms of these
\be\label{reninva}
a=\alpha_1A-\alpha_2A^3+{\tt o}(A^5)\,,\quad
g=-1+\alpha_1^2A^2-{1\over2}\alpha_1^4A^4+{\tt o}(A^6)\,.
\ee

\subsection{Unified description}

In order to make the computations more general it is useful to express
both theories as particular cases of a generalized \szm . To achieve this
we rescale the field $\psi\longrightarrow \frac{1}{\sqrt{1+g}}\psi$
and introduce $a=\sqrt{1+g}$ in place of $g$ in the original model. After
some obvious changes of symbols, ${\cal L}_O$ and ${\cal L}_D$ can be
described by the following unified \szm :
\be\label{lagrsigunif}
\eqalign{
{\cal L}=&{1\over2\lambda}\bigl\{ (\pa_\mu\theta)^2+(\pa_\mu\phi)^2\bigl(
1+r
\cos^2\theta\bigr)+
(\pa_\mu\psi)^2+2a\omega_{\mu\nu}\cos\theta\pa_\mu\phi\pa^\mu\psi\bigr\}\,.
}\ee
The previous models can be recovered by the special choices: $r=g$ and
$\omega_{\mu\nu}=\eta_{\mu\nu}$ for the original model, and $r=-1$ and
$\omega_{\mu\nu}=-\epsilon_{\mu\nu}$ for the dual model. 
Expressing $r$ in terms of the RG invariant parameter we have:
\be\label{fundpar1}
r=-1+2\beta A^2+\gamma A^4+{\tt o}(A^6)\,,
\ee
where $\beta=\alpha_1^2/2$ and $\gamma=-{1\over4}\alpha_1^4$ in the 
original resp. $\beta=0$ $\gamma=0$ in the dual model. Observe that both
$\beta$ and $\gamma$ are renormalization group (RG) invariant.

This unified Lagrangian (\ref{lagrsigunif}) is more than it might appear
at first sight. As it was shown in \cite{bfhp} and argued above, the
deformed principal \szm \ and its dual can be viewed as being quantum
equivalent from the point of view of the two loop beta functions. The
unifying Lagrangian (\ref{lagrsigunif})  can be viewed as a genuine
quantum generalization of the deformed principal \szm \ and its dual. It
reduces to the latter ones at special values of the parameters
$\beta,\,\gamma$ and $\omega_{\mu\nu}$, and at the same time gives the
corresponding beta functions. Thus the renormalization properties of the
two models are encompassed in this generalized Lagrangian.
Based on this we will be able to renormalize the free energy in both
theories at the same time, shortening the computations and obtaining
better control on the different contributions.. 
Checking the quantum equivalence will amount to compare
the  renormalized free energies computed at the special values of the
parameters $\beta,\,\gamma$ and $\omega_{\mu\nu}$ corresponding to the
two models.

\section{The ground state and perturbative Lagrangian}

\subsection{Outline of the method - Noether currents}

So far testing the quantum equivalence of the dual models was mostly
reduced to comparisons of the corresponding beta-functions
\cite{bfhp,en,haagensen1}.  Of course there are more to test before one
can ascertain about an equivalence. In this paper to test the physical
equivalence between the original and dual theories we couple both of them
to some particular conserved charge $Q$. This is accomplished by modifying
the respective Hamiltonians $H_O$ $(H_D)$ to $H_O-hQ$ $(H_D-hQ)$, where
$h$ is an external field (chemical potential type of parameter), having
mass dimension one. The corresponding changes in the ground state energy
densities (i.e. in the free energy densities, that we shall call for
simplicity free energies) can be computed, at least in principle, to any
order in perturbation theory. The comparison of the free energy densities,
as functions of $h$, up to a certain order of perturbation theory, in the
original and dual models, then provides a useful check whether the two
models do really, physically correspond to each other. For a comparison of
this type to make sense both theories must be asymptotically free (to
guarantee that perturbation theory applies), and of course we have to
choose $Q$ to be really the same.

As the global symmetry group of both the original and the dual models is
$SU(2)\times U(1)$ we may think naively that any linear combination of an
$SU(2)$ Noether charge and the $U(1)$ Noether charge, $x^a Q^a+y
Q_{U(1)}$, can be used as the charge above $Q$ to couple to the
Hamiltonians. However for our test we need the same $Q$ coupled to $H_O$
and $H_D$, thus we can choose only such charges that are mapped to
themselves under the canonical transformation connecting the original and
dual models. This, of course, implies that the charge $Q$ must stay local
under the canonical transformation.

We first point out the relation between the appropriate Noether currents
and charges of the global symmetries of ${\cal L}_O$ and ${\cal L}_D$. An
exhaustive treatment will be given elsewhere \cite{en2}. Let's start
remarking that under Abelian duality transformations the image of the
$U(1)$ current, that belongs to the distinguished isometry used in
duality, is a {\em topological current} built from the dual field. Thus
the image of the $U(1)_R$ current of the deformed sigma model is the
topological current of the $\chi$ field, 
$\epsilon_{\mu\nu}\partial^\nu \chi$, and as such, its
charge should vanish on a topologically trivial 2d space-time.  Therefore
$Q_{U(1)}$ cannot be present in both $H_O$ and $H_D$.

The next simplest possibility is to use the $U(1)$ charges corresponding
to the $\phi$-translations in both the original and dual theories:  
${N}^3_0$ and $\tilde{N}^3_0$. It can be shown (see \cite{en2} for
details) that the canonical transformation which implements Abelian
duality, effectively exchanges only $p_\psi$ and $\chi '$ ($p_\chi$ and $\psi '$)
and leaves  ${p}_\phi$, ${p}_\theta$ unchanged. It is obvious that $\tilde{N}^3_0$ is
'identical' under duality to $N^3_0$, but it is not entirely trivial
(though it is true) that the space component of $N^3_\mu$ really becomes
the space component of $\tilde{N}^3_\mu$.

\subsection{From the Hamiltonian to the perturbative Lagrangian}
    
In conclusion, to test the physical equivalence of the original and dual
models, in both of them we introduce the coupled Hamiltonian densities
$\bar{\cal H}_{O,D} ={\cal H}_{O,D} -hN_0^3$.\footnote{One can also
investigate different linear combinations of the Noether charges
\cite{en2}.} Performing the inverse Legendre transformation on these
quantities one obtains the $\bar{\cal L}_{O,D}$ Lagrange densities of the
coupled models. By explicit computation one can show that this procedure
of obtaining $\bar{\cal L}_{O,D}$ is equivalent to the following formal
gauging ($\pa_\mu\zeta^i\rightarrow D_\mu\zeta^i$) of the original
Lagrangian ${\cal L}_{O,D}$: 
\be\label{gauge} 
D_\mu\psi=\pa_\mu\psi,\quad
D_\mu\phi=\pa_\mu\phi+h\delta_{0\mu},\quad D_\mu\theta=\pa_\mu\theta. 
\ee
This gauging, of course, can be done on the universal, common Lagrangian
(\ref{lagrsigunif}); the outcome being denoted by $\bar{\cal L}$. As it
can be seen in (\ref{gauge}), the coupling of the chemical potential
explicitly breaks Lorentz invariance. This will play an important role in
our analysis of the quantum equivalence.

%\subsection{The ground states of $\bar{\cal H}$}

To perform the perturbative analysis first one has to determine the
classical ground state of the system, in other words the minima of the
Hamiltonian $\bar{\cal H}=\bar{\cal H}
(\theta,\phi,\psi,p_\theta,p_\phi,p_\psi)$, finding the critical points
and checking the Hessian's positive definiteness. Using the actual
(perturbative) expression for r, eq. (\ref{fundpar1}), it is
straightforward to show that in the perturbative region around $r=-1$
(which in the original model corresponds to $g=-1$ as opposed to the dual
model where it is exactly -1) there is a two-parameter family of local
minima, all of them being physically equivalent, given by:
$(p_\theta=p_\psi=0,p_\phi=\frac{h}{\lambda},\theta=\frac{\pi}{2},
\phi=const.,\psi=const.)$. Since $\phi$ and $\psi$ does not appear
explicitly in the Hamiltonian, we can chose for convenience the stable
classical ground state, common to both models, to be given by
$\theta={\pi\over 2},\, \phi=\psi=0$. We shall expand our fields around
this background.

%\subsection{Rescaling of the fields}

The Lagrangian, $\bar{\cal L}$, as emerging from eq. (\ref{lagrsigunif})  
is not suitable for a direct perturbative computation on account of the
overall $1/\lambda$ factor. However this factor can be removed by
an appropriate rescaling of the fields:
\be
\label{rescale}
\tilde{\theta}=\frac{\theta}{A},\qquad\tilde{\phi}=\frac{\phi}{A},\qquad
\tilde{\psi}=\frac{\psi}{A},
\ee
where $A=\sqrt{\lambda}$ as it was given in (\ref{fundpar}). We note here
that originally the fields had no  
renormalization in either the original or in the dual model, consistent
with the fact that the Euler angles are compact variables. The
rescaling nevertheless introduces nontrivial renormalization, but this can
be deduced from the renormalization of $\lambda$.

Deleting the tilde from the rescaled fields, we get the following result:
\be\label{pertlagsub}
\eqalign{
\bar{\cal L}=&{1\over2}\bigl\{ 
(\pa_\mu\theta)^2+\bigl( 
1+r\cos^2(A\theta)\bigr)(\pa_\mu\phi+\frac{h}{A}\delta_{0\mu})^2+\cr
&+(\pa_\mu\psi)^2+2a\omega_{\mu\nu}\cos(A\theta) 
(\pa_\mu\phi+\frac{h}{A}\delta_{0\mu})\pa_\nu\psi\bigr\}\,. 
}\ee

%\subsection{Euclidian continuation}

According to the general prescription of Wick rotation, we define the
Euclidian continuation of our model by $\bar{\cal L}_E=-\bar{\cal
L}_M(\pa_0\rightarrow i\pa_0,\pa_1\rightarrow\pa_1)$. Note that a similar
continuation of $h\rightarrow ih$ would lead to inconsistencies (like
wrong signs in the propagators: $\frac{1}{p_E^2-h^2}$ instead
of$\frac{1}{p_E^2+h^2}$).

Before starting the perturbative expansion of the trigonometric functions
we make another field transformation (redefinition):  
$\theta\rightarrow\frac{\pi}{2\lambda}+\theta$, guaranteeing that the
minimum we expand around is: $(\theta=0,\phi=0,\psi=0)$. Thus the
Euclidian Lagrangian we use has the form:  
\be\label{pertlagsub1}
\eqalign{ 
\bar{\cal L}=&{1\over2}\bigl\{ (\pa_\mu\theta)^2+\bigl(
1+r\sin^2(A\theta)\bigr)(\pa_\mu\phi+\frac{h}{A}\delta_{0\mu})^2+\cr
&+(\pa_\mu\psi)^2+2a\sin(A\theta)
(\omega_{\mu\nu}\pa_\mu\phi\pa_\nu\psi+\frac{h}{A}
{\Omega}_{0\nu})\pa_\nu\psi\bigr\}\,, 
} 
\ee 
where $r$ and $a$ are
given by (\ref{fundpar1}) resp. (\ref{reninva}), and the new parameters
$\omega$ and $\Omega$, which bear the model dependence, are:  
$\omega_{\mu\nu}=-\delta_{\mu\nu}$, ${\Omega}_{\nu}=i\delta_{0\nu}$ in the
original and
$\omega_{\mu\nu}=-i\epsilon_{\mu\nu}$, ${\Omega}_{\nu}=-\epsilon_{0\nu} $
respectively in the dual model.

%\subsection{The perturbative Lagrangian}

Next we expand the Lagrangian $\bar{\cal L}$ around the classical ground
state: $(\theta=0,\phi=0,\psi=0)$, with $A=\sqrt{\lambda}$ being  
 the relevant coupling constant.  We note at this point that one could
follow a different route and expand the parameters $r$, $ g$ and $a$ in
terms of the RG invariant. The motivation for this would be that at the
end of the computation this has to be done anyway. This possibility and
the complications that arise will be investigated elsewhere \cite{en2}.

After some algebra the result is as follows (for the sake of
simplicity
we denote $\bar{\cal L}$ by ${\cal L}$):
\be\label{lpertot}
{\cal L}={\cal L}_{-2}A^{-2}+{\cal L}_{-1}A^{-1}+{\cal L}_{0}+
+{\cal L}_{1}A+{\cal L}_{2}A^{2}+{\tt o}(A^3)\,,
\ee
where
\be\label{lpert}
\eqalign{
{\cal L}_{-2}=&-\frac{1}{2}h^2\,,\qquad\qquad 
{\cal L}_{-1}=-ih\,\pa_0\phi\,,\cr
{\cal L}_{0}=&\frac{1}{2}(\pa_\mu\psi)^2+ \frac{1}{2}(\pa_\mu\phi)^2+ 
\frac{1}{2}(\pa_\mu\theta)^2-r\frac{1}{2}h^2\theta^2+
ah{\Omega}_{\nu}\,\theta\pa_\nu\psi\,,\cr
{\cal
L}_{1}=&-irh\,\theta^2\pa_0\psi+a\omega_{\mu\nu}\,\theta\pa_\mu\phi\pa_\nu\psi
\,,\cr
{\cal L}_{2}=&\frac{1}{2}r\theta^2(\pa_\mu\phi)^2+\frac{1}{6}rh^2\theta^4
-\frac{1}{6}ah{\Omega}_{\nu}\theta^3\pa_\nu\psi \,.
}\ee
Notice that ${\cal L}_{-2}$ is a constant (i.e., it is independent of the
fields), while ${\cal L}_{-1}(h,\varphi)=-ih\,\pa_0\phi$ is a total
derivative, thus it can be discarded in this non-topological sector of the
theory. From ${\cal L}_0$ we see that $\phi$ is a massless scalar field,
while $\theta$ and $\psi$ are mixed, apart of the mixing the former is a
massive scalar field with mass $\sqrt{-r}h$, the latter is massless. The
interaction of the different fields is highly non-trivial, as can be seen
above, and contains infinitely many vertices. Nevertheless, these vertices
are naturally separated in the weak coupling regime.  We emphasize that
only the first two terms, ${\cal L}_{-2}$ and ${\cal L}_{-1}$, are common
to both models, as the model dependent parameters $\alpha_i$, $\beta$,
$\omega$ and $\Omega$ appear in ${\cal L}_j$ for all $j\geq 0$.

\section{The free energy}

Our goal is to define and compute the free energy density in perturbation
theory. After setting the stage we do the explicit
computations.

\subsection{Definition of the free energy}

At this point our aim is to define the free energy density
perturbatively. Denoting the fields collectively by
$\varphi=(\theta,\phi,\psi)$, the free energy (density) reads:
\be\label{freen}
e^{-{\cal F}(h)\,V}=\frac{\int\! {\cal D}\varphi\,
e^{-S[h,\varphi]}}{\int\!
{\cal D}\varphi\, e^{-S[h=0,\varphi]}}\,,
\ee
where $V$ is the volume of the system, $S[h,\varphi]= \int\! d^2x\, {\cal
L}(h,\varphi(x))$, and $\int\! {\cal D}\varphi$ denotes the functional
integration over the field configurations $\varphi=(\theta,\phi,\psi)$: 
$\int\! {\cal D}\varphi=\int\! {\cal
D}\phi {\cal D}\psi{\cal D}\theta$.
The role of the denominator in (\ref{freen}) is to insure the correct
normalization: ${\cal F}(h=0)=0$. From dimensional arguments one expects
the following functional dependence: ${\cal F}(h)=h^2\Psi(h)$, where
$\Psi=\Psi(h)$ is a dimensionless function.

Let us note the similarity between the free energy defined above and
quantum effective action (the generator of the 1PI graphs). The role of
the external field is played by $h$, that couples to a conserved charge
(composite operator)
rather than an elementary field. This similarity will play a 
structurally simplifying role when
we discuss the renormalization of the model.

For the perturbative expansion, in view of eqs.
(\ref{lpertot},\ref{freen}), it proves useful to introduce
\be\label{freenex}
S[h,\varphi]= \sum_{i=-2}^\infty S_i[h,\varphi]\,A^i\,,
\ee
with $S_i[h,\varphi]= \int\! d^2x \,{\cal
L}_i(h,\varphi(x))$, where $i\geq -2$. Using this in eq. (\ref{freenex})
we obtain a similar expression for the free energy:
\be\label{fre}
{\cal F}(h)= \sum_{i=-2}^\infty {\cal F}_i(h)\,A^i\,.
\ee
Our task will be to determine the first few terms in this expansion, in
both models, and compare them. More precisely we determine the first six
terms of ${\cal F}(h)$  and check whether they are equal.

As ${\cal L}_{-2}$ is independent of the fields $\varphi$, it results
that
$S_{-2}[h,\varphi]=-\frac{1}{2}h^2V$, and $\exp (-S_{-2}[h,\varphi])$
factorizes (we will come back to this) in the functional integral
(\ref{freen}). This way one readily
obtains that the first term of (\ref{fre}) is
\be\label{frem2}
{\cal F}_{-2}(h)=-\frac{1}{2}h^2\,.
\ee
Of course this is valid both in the original and dual model, implying that
at leading order the perturbative free energy densities coincide.  In
addition, since ${\cal L}_{-1}$ is a total derivative, it implies that
$S_{-1}[h,\varphi]=0$, thus
\be\label{frem1}
{\cal F}_{-1}(h)=0\,;
\ee
again a model independent statement.

Thus all what remains to be dealt with is the reduced action
\be\label{freenexr}
\bar{S}[h,\varphi]= \sum_{i=0}^\infty S_i[h,\varphi]\,A^i\,,
\ee 
and the reduced free energy
\be\label{frer}
\bar{\cal F}(h)= \sum_{i=0}^\infty {\cal F}_i(h)\,A^i\,,\quad
{\cal F}(h)={\cal F}_{-2}(h)+\bar{\cal F}(h)\,.
\ee
Eq. (\ref{frem1}) might also suggest that all ${\cal F}_{i}$ with $i$ odd
vanishes, this would comply with the fact that the coupling
$A=\sqrt{\lambda}$ is just an artifact of our perturbation theory,
as it was ${\lambda}$ that appeared in the original Lagrangian. The
vanishing of all odd power contributions in $A$ would indeed imply that 
the true
coupling is in fact $\lambda$.

Introducing the $Z(h)=\int\! {\cal D}\varphi\, e^{-\bar{S}[h,\varphi]}$
auxiliary function then we can rewrite eq. (\ref{freen}) as $e^{-\bar{\cal
F}(h)V}= Z(h)/Z(h=0)$. Moreover, if $M$ is an operator we define the
following 'expectation value':
\be\label{average}
\langle M\rangle =\frac{\int\! {\cal D}\varphi\, e^{-S_0[h,\varphi]}\,M}{
\int\!
{\cal D}\varphi\, e^{-S_0[h,\varphi]}}\,.\ee
Expanding $Z(h)$ as a power series in $A$ we have: 
\be\label{inerm}
\eqalign{
Z(h)=&\big[\,1-\langle S_1\rangle  A-\langle
S_2-{1\over2}S_1^{\,2}\rangle A^2-\cr
&-\langle S_3-S_1S_2-{1\over6} 
S_1^{\,3}\rangle  A^3+ \vartheta (A^4)\,\big]\int\! {\cal D}\varphi\,
e^{-\bar{S}[h,\varphi]}\,.
}
\ee
For simplicity we have
omitted to write the functional dependence of the
$\bar{S}_i[h,\varphi]$-s.
Using the identity
\be\label{ident}
1+y_1\,A+y_2\,A^2+y_3\,A^3+ {\tt o} (A^4)=e^{
x_1\,A+x_2\,A^2+x_3\,A^3+ {\tt o} (A^4)}\,,
\ee
where
\be\nonumber
x_1=y_1\,,\quad x_2=y_2-{1\over2}y_1^{\,2}\,,\quad x_3=y_3-y_1\,y_2
+{1\over3}y_1^{\,3}\,,
\ee
we can read off the various components of the reduced free energy density
\be\label{freen0}
e^{-{\cal F}_0(h)\,V}=\frac{\int\! {\cal D}\varphi\,
e^{-S_0[h,\varphi]}}{\int\! {\cal D}\varphi\, e^{-S_0[h=0,\varphi]}}\,,
\ee
\be\label{freen1}
{\cal F}_1(h)=\frac{1}{V}\langle S_1\rangle \,,
\ee
\be\label{freen2}
{\cal F}_2(h)=\frac{1}{V}\,\big[\langle S_2-{1\over2}S_1^{\,2}\rangle + 
{1\over2}\langle S_1\rangle ^2\big]\,.
\ee
We emphasize that in the above formulas we kept only those terms which
depend on $h$. In other words we discarded the contribution of $Z(h=0)$,
which in fact is a divergent quantity. This is consistent with the fact
the that only the derivatives of the free energy are observable 
thermodynamical quantities.

\subsection{Propagators}

To compute the various vacuum expectation values (or correlators since we
are in Euclidean space) determining ${\cal F}_i(h)$ we use dimensional
regularization (with $n=2+\epsilon$). 
Since
\be\label{act0}
S_0(h)=\int\! d^2x \,{\cal L}_{0}(h)=\frac{1}{2}\int\! d^2x \,\varphi^t(x)
M(x)\varphi (x)
\ee
where
\be\label{act01}
\eqalign{
&\varphi^t (x)=(\theta  (x),\phi (x),\psi (x))\,,\quad\cr
&M(x)=\left(
\begin{array}{ccc}
-\pa^2-rh^2&0&ah\Omega_\nu\pa_\nu\\
0&-\pa^2&0\\
-ah\Omega_\nu\pa_\nu&0&-\pa^2\\
\end{array}\right)
\,,\quad
\pa^2=\pa_\mu\pa_\mu\,,
}
\ee
in order to determine the propagators of the various fields we have to
invert the matrix operator $M(x)$. This is easily done in momentum space
resulting:
\be\label{propag}
\begin{array}{ll}
{G}_{\theta}(x)=\int\!\frac{d^np}{(2\pi)^n}\,e^{-ipx}\,p^2g(p)\,,&
{G}_{\psi}(x)=\int\!\frac{d^np}{(2\pi)^n}\,e^{-ipx}\,(p^2-rh^2)g(p)\,,\\
{G}_{\theta\psi}(x)=iah\!\int\!\frac{d^np}{(2\pi)^n}\,e^{-ipx} 
\Omega\!\cdot\! p\,g(p)
,&{G}_{\phi\phi}(x)=\int\!\frac{d^np}{(2\pi)^n}\,e^{-ipx}\, 
\frac{1}{p^2}\,,\\
{G}_{\theta\phi}(x)=0,&{G}_{\psi\phi}(x)=0,\\
\end{array}
\ee
where
\be\label{g}
g(p)=\frac{1}{p^4-h^2[rp^2+(a\Omega\cdot p)^2]}\,, \quad \Omega\cdot
p\equiv\Omega_\nu   
p_\nu.
\ee
We have used commonly the notation ${G}_{\varphi\varphi'}(x-y)=
\langle\varphi(x)\varphi'(y)\rangle$, with
${G}_{\varphi}\equiv{G}_{\varphi\varphi}$.  
Just note in passing a few simple properties:
${G}_{\theta\theta}(x)$ is even, while ${G}_{\theta\psi}(x)$ and
${G}_{\psi\psi}(x)$ are
odd, and ${G}_{\psi\theta}(x)=-{G}_{\theta\psi}(x)$. It is also obvious
that $\phi$ behaves like a massless scalar.

\subsection{Computation of ${\cal F}_0(h)$}

The computation of ${\cal F}_0(h)$ involves in fact the evaluation of a
functional determinant, similarly to the the case of the quantum effective
action.
Evaluating the Gaussian integral from eq. (\ref{freen0}), 
using the
identity $\det X=e^{{\rm Tr\,}\ln X}$, results:
\be\label{f0}
{\cal F}_0(h)=\frac{1}{2}\int\! \frac{d^np}{(2\pi)^n}\,\ln \big(1-
\frac{h^2(rp^2+(a\Omega\cdot p)^2)}{p^4}\big)\,.
\ee
A proper way to compute this expression is to take its derivative with
respect to $h$ and solve the following initial value problem:
\be
\frac{d{\cal
F}_0(h)}{dh}=-h\,\int\!\frac{d^np}{(2\pi)^n}\,\frac{rp^2+(a\Omega\cdot
p)^2}{p^4-h^2(rp^2+(a\Omega\cdot
p)^2)}
\,,\quad {\cal F}_0(0)=0\,.
\ee
Rescaling $p\rightarrow hp$, the $h$ dependence factorizes and we get:
\be\label{fre0}
{\cal
F}_0(h)=-\frac{h^{n}}{n}\,\int\!\frac{d^np}{(2\pi)^n}\,\frac{r_0p^2+
(a_0\Omega\cdot p)^2}{p^4-h^2[r_0p^2+(a_0\Omega\cdot p)^2]} \,. 
\ee
Above we have made it explicit that the integral is computed in terms of
the bare quantities $r_0$ and $a_0$, rather then the renormalized ones,
$r$ and $a$. From now on we are going to make this distinction clear in all 
subsequent formulas.

This same expression was obtained in \cite{pall2}, though the initial
Lagrangian differed from the one used here by a certain rescaling of the
fields. A rescaling usually cannot cause major discrepancies, at least at
low orders of perturbation theory, as this example also reflects.

The model dependence is manifest in (\ref{fre0}). The integral is
divergent in two dimensions more precisely it has a first order pole in
$\epsilon$. The analysis of \cite{pall2} has computed the pole term and
the constant term in the $\epsilon$ expansion of (\ref{fre0}) in closed
form, in terms of generalized hyper-geometric functions. After
renormalization, the two
expressions -- as functions of the original (respectively dual) coupling
and parameter -- 
were not equal, but, the difference could be accounted for by the
scheme dependence of the two loop beta functions. Indeed it was  pointed
out in \cite{bfhp} that the equivalence of the original and dual $\beta$ 
functions corresponds to a perturbative redefinition of the
coupling constants in the dual model:
\be\label{psicoupl}
\tilde\lambda=\lambda+{\lambda^2\over4\pi}(1+g)\,,\quad
\tilde g=g+{\lambda\over4\pi}(1+g)^2\,.
\ee
Implementing this redefinition in the expressions of the renormalized
one loop free energy densities revealed their equality.  

Here we take a different route from the one described above. Our strategy
is to express the parameters that bear the model dependence in terms of
the RG invariant in both models, then set these two RG invariants equal
and compare the results. Evaluating $-r_0p^2- (a_0\Omega\cdot p)^2$ we get
$p_0^2-g_0p_1^2$ in the original model and $\tilde{p}_1^2-g_0\tilde{p}_0^2$
respectively in the dual model, where
$\tilde{p}_\mu=\epsilon_{\mu\nu}p_\nu$. Due to \cite{epsilon} we have
$\tilde{p}^2=p^2$, and one can perform a change of variables from $p$ to
$\tilde p$. This way, changing also $p_1\leftrightarrow p_0$, one can
obtain formally identical expressions in the two cases. At this point we
only remark that the role of $p_0$ and $p_1$ in dimensional regularization
is different.
   
\subsection{Computation of ${\cal F}_1(h)$}
 
According to (\ref{freen1}) and (\ref{lpert}) the computation of ${\cal
F}_1(h)$ involves the following correlation functions: $\langle\theta
(x)\pa_\nu\psi (x)\rangle$ and $\langle\theta^2(x)\pa_0\phi(x)\rangle$.
Using Wick's theorem
we obtain:
\be\label{fre1}
{\cal F}_1(h)=0\,.
\ee
Eq. (\ref{fre1}) has twofold meaning. 
It shows once again 
the model independence of the free energy, though we have already 
encountered explicit model dependence. On the other hand 
supports our
earlier statement about the vanishing of the non-analytic corrections in
$\lambda$. 

\subsection{Computation of ${\cal F}_2(h)$}

In the case of ${\cal F}_2(h)$ we will not be able to obtain the result in
a closed form, nevertheless what we can actually compute will suffice to
achieve our goals. According to eq. (\ref{freen2}) we have to compute
$\langle S_2(h)\rangle$, $\langle S_1(h)^2\rangle$ and $\langle
S_1(h)\rangle^{\,2}$. We can immediately quote eq. (\ref{fre1}) and
$\langle S_1(h)\rangle=0$.

\subsubsection{Computation of $\langle S_2(h)\rangle$}

From eq. (\ref{lpert}) we see that
\be\label{s2}
\eqalign{
\langle S_2(h)\rangle=\int\!  d^nx&
\,[\frac{1}{2}r_0\langle \theta^2(x)(\pa_\mu\phi)^2(x)\rangle +
\frac{1}{6}r_0h^2 \langle \theta^4(x)\rangle-\cr &
-\frac{1}{6}a_0h\Omega_\nu\langle \theta^3\pa_\nu\psi(x)\rangle]\,.  
}
\ee
At first sight one might want to discard the terms that are not coupled 
to $h$. 
Nevertheless these terms acquire $h$ dependence through the $h$-dependent 
propagators.

Based on Wick's theorem for the first term we have $\langle
\theta^2(x)(\pa_\mu\phi)^2(x)\rangle =\langle \theta^2(x)\rangle \langle
(\pa_\mu\phi)^2(x)\rangle $.  Since the $\phi$ propagator is in fact a
Green's function, or in other words fundamental solutions of the
corresponding wave equation, we can conclude that $\langle
(\pa_\mu\phi)^2(x)\rangle =\delta^{(n)}(0)$ (modulo equal time commutator
terms), where $\delta^{(n)}(0)$ is the Dirac delta distribution 'evaluated
at 0'. But in dimensional regularization the latter is set to zero,
yielding no contribution.

The next term in eq. (\ref{s2}) is also readily evaluated: $\langle
\theta^4(x)\rangle =3\langle
\theta^2(x)\rangle ^2=3\,{G}_{\theta}(0)^2$. 
Evaluating the third term
gives (recall that the system in finite volume: $\int\!
d^nx=V$):
\be\label{s2f}
\langle S_2(h)\rangle ={1\over 
2}h^2V\,\left[r_0+\frac{1}{n}a_0^{\,2}\Omega_\nu\Omega_\nu\right]
{G}_{\theta}(0)^2 \,.
\ee

\subsubsection{Computation of $\langle S_1(h)^2\rangle $}

The computation of $\langle S_1(h)^2\rangle $ involves in fact the
evaluation of double integral $\int\!
d^nx \,d^ny \,\langle {\cal L} (x){\cal L} (y) \rangle$. From eq.
(\ref{lpert}) it turns out that $\langle S_1(h)^2\rangle $ equals
$\int\! d^nx$ times
\be\label{s12}
\eqalign{
&r_0^2h^2\pa_0^{\,2}G_\phi(x)[G_\theta(0)^2+2G_\theta(x)^2]
-4ir_0ha_0\omega_{\mu\nu}\pa_0\pa_\mu G_\phi(x)G_\theta(x)\pa_\nu 
G_{\theta\phi}(x)+ \cr
&+\quad a_0^{\,2}\omega_{\mu\nu}\omega_{\lambda\rho}\pa_\mu\pa_\lambda 
G_\phi(x)[-\pa_\nu G_{\theta\phi}(x)\pa_\rho G_{\theta\phi}(x)
-G_\theta(x)\pa_\nu\pa_\rho G_{\psi}(x)+\cr
&\quad +\pa_\nu G_{\theta\phi}(0)\pa_\rho 
G_{\theta\phi}(0)]\,.
}
\ee

A priori it is not clear at all how to obtain the overall $h^2$ factor
required by dimensional analysis. Moreover it is also puzzling how the
explicit factors of $i$ disappear during the computation. As we shall see
shortly it is the form of the \lq tensor' parameters $\Omega_\nu$, $\omega_{\mu\nu}$
and propagators that is responsible for the correct answers.

The first term in (\ref{s12}) 
results
\be\label{f21}
-2r_0^{\,2}h^2V{1\over n}G_{\theta}(0)^2\delta_{00}\,.
\ee
The only non-trivial fact that one has to use an IR regularization of the
field $\phi$, the regulator's mass we denote by $m$. Then the first
term in this parenthesis will be proportional to 
\be\label{int2}
\int\! d^nx\,\pa_0^{\,2} G_\phi(x)\approx \int\!
d^nk\frac{k_0^{\,2}}{k^2+m^2}\, 
\delta^{(n)}(k)=0\,.
\ee

The second term  in (\ref{s12})  results
\be\label{f22}
4r_0a_0^{\,2}h^2Vi\omega_{\mu\nu}\Omega_\rho
\int\!\frac{d^nk_1}{(2\pi)^n}\frac{d^nk_2}{(2\pi)^n}
\frac{(k_1+k_2)_0(k_1+k_2)_\mu}{(k_1+k_2)^2}k_1^{\,2}g(k_1)k_{2\nu} 
k_{2\rho}g(k_2)\,,
\ee
while the last terms  of (\ref{s12})  give
\be\label{f23}
\eqalign{
a_0^{\,2}\omega_{\mu\nu}\omega_{\lambda\rho}V
\int\!\frac{d^nk_1}{(2\pi)^n}\frac{d^nk_2}{(2\pi)^n}&
\frac{(k_1+k_2)_\lambda(k_1+k_2)_\mu}{(k_1+k_2)^2}g(k_1)g(k_2)k_{2\rho}
\cr
&[a_0^{\,2}h^2k_{1\nu}(k_1\cdot\Omega)(k_2\cdot\Omega)+k_1^{\,2}k_{2\nu} 
(k_2^{\,2}-r_0h^2)] \, . 
}
\ee

The general structure of the integrals that appear (with one
exception) is of the following
form:
\be\label{gf}
\int\!\frac{d^nk_1}{(2\pi)^n}\frac{d^nk_2}{(2\pi)^n}
\frac{(k_1+k_2)_{\mu_1}(k_1+k_2)_{\mu_2}}{(k_1+k_2)^2}k_{1_{\mu_3}}
k_{1_{\mu_4}}k_{2_{\mu_5}}k_{2_{\mu_6}}g(k_1)g(k_2).
\ee
The only integral that cannot be brought to this form has $k_2^2g(k_2)$
instead of $g(k_2)$. If $g(k)$ were a covariant expression in $k$, then
the value of the integral would be given completely by the index
structure. More precisely covariance would require the result to be the
sum of triple products of the
Kronecker delta functions. The number of independent possibilities is
$6!/(2!)^3/3!=15$. Exploiting the obvious symmetries under the exchange of
$\mu_1\leftrightarrow\mu_2$, $\mu_3\leftrightarrow\mu_4$,
$\mu_5\leftrightarrow\mu_6$, and the less obvious exchange
$(\mu_3,\mu_4)\leftrightarrow(\mu_5,\mu_6)$, which can be seen by a change 
of integration variables, the tensor structure reduces significantly to
only four
independent terms. Denoting $\delta_{\mu_i\mu_j}$ by $(ij)$, we get for
(\ref{gf})
\be\label{dr}
\eqalign{
&(12)(34)(56)\,I_1+(12)[(35)(46)+(36)(45)]\,I_2+(34)[(15)(26)+(16)(25)]\,I_3+
\cr
&+(56)[(13)(24)+(14)(23)]\,I_3+\{(13)[(25)(46)+(26)(45)]+(14)[(25)(36)+
\cr
&+(26)(35)]+(15)[(24)(36)+(23)(46)]+(16)[(23)(45)+(24)(35)]\}\,I_4.
}
\ee
Using the standard methodology one can compute the unknowns $I_1$ through
$I_4$, in a straightforward manner. Unfortunately $g(k)$ is not covariant
with respect to the full $SO(n)$. As we pointed out, covariance was
already
broken at the level of the Lagrangian. If we analytically continue
$p_0$ to $n_0$ dimensions and $p_1$ to $n_1$ dimensions, with $n_0+n_1=n$,
then instead of the full $SO(n)$ group we get $SO(n_0)\times SO(n_1)$. In
other words, among the $p_0$-s and $p_1$-s standard covariance arguments
remain valid, and the computation sketched above makes sense. Thus if all
the indices are of $p_0$ or $p_1$ type, then we can reliably compute the
integrals.

In the expression of the two-loop free energy (\ref{f22}, \ref{f23}) the
free indices we have dealt above are contracted with different tensor
structures. In the original model $\Omega_{\nu}$ and $\omega_{\mu\nu}$
involve only delta functions, and this way what we said applies. On the
other hand for the dual model $\Omega_{\nu}$ and $\omega_{\mu\nu}$ are
epsilon tensors, mixing the indices, and the arguments presented break
down. The same is true for the sum of the terms involved, that we
eventually want to compute.

There is an independent argument that shows that even if we were able to
do the integrals, the result would not be reliable, due to the behavior of
the $\epsilon$ tensor in dimensional regularization. More precisely, even
if the integrals in (\ref{f22}) and (\ref{f23}) were covariant, based on
the definition of $\omega_{\mu\nu}$ and on (\ref{dr}), we could conclude
that in the dual model $\langle S_1(h)^2\rangle $, and as a result ${\cal
F}_2(h)$, contains terms proportional to the product of two $\epsilon$
tensors, with uncontracted indices. The broken covariance by the external
field $h$, invalidates the above argument, but is highly likely that it
complicates matters, rather than simplifies, and as a result we would
still end up with $\epsilon_{\mu\nu}\epsilon_{\alpha\beta}$ terms. As it
is well known, there is no consistent way defining such on object in
dimensional regularization \cite{epsilon}.

At this point the computation of $\langle S_1(h)^2\rangle $ is hopeless,
driving the same statement about ${\cal F}_2(h)$. Nevertheless we can do
something less ambitious. Following \cite{bfhp,en}, in order to compare
the results, at the very end of the computation we want to express them in
terms of the two RG invariants, which are finally set equal. In other
words, based on (\ref{reninv1}) and (\ref{fundpar1}) we trade the
renormalized quantities $g$ and $r$ for the RG invariant $M$, implicitly
meaning that by this we have also set the RG invariants equal. Since it is
the bare $a_0$ that appears in (\ref{f22}) and (\ref{f23}), and also in
(\ref{fre0}), let's investigate more closely what happens to this term
during renormalization and expansion in terms of the RG invariant.

At the level of the bare quantities we aim to have $a_0=\sqrt{1+ g_0}$.
Based on \cite{bfhp} this is certainly true at two loop level in the
original model, while in the dual one this is more subtle. As it was shown
in \cite{bfhp} naively the relation $a_0=\sqrt{1+ g_0}$ cannot be
maintained at two loop level in the dual model. Nevertheless taking into
account the redefinition of the dual model's coupling and parameter,
eq.(\ref{psicoupl}), the above relation can be maintained at one loop
order.

As $a_0=\sqrt{1+ g_0}$ is doubtlessly valid in the leading order, the
renormalization (\ref{couplren1})  amounts to $a_0=\sqrt{1+
Z_g(\lambda,g)g}$. But (\ref{bf1}) shows that $Z_g(\lambda,g)=1+{\tt
o}(\lambda)$, thus to leading order we have $a_0=\sqrt{1+ g}$, with $g$
the renormalized coupling. On the other hand at this point we can use
(\ref{reninv1}) and conclude that, after renormalization and expanding in
the RG invariant, $a_0$ becomes proportional to $\sqrt{\lambda}$.

The good news is that all the terms we were unable to compute (\ref{f22})
and (\ref{f23}) are proportional to $a^2$, hence are of order $\lambda^2$.
Since we have not even attempted to compute the ${\cal F}_4(h)$ term that
is of the same order, we neglect them for the time being. Having in mind
the insertion of the RG
expressions for the parameters, we simply get: 
\be\label{fre2} 
{\cal F}_2(h)=-h^2\frac{n-2}{2n} G_{\theta}(0)^2\,.  
\ee 
Naturally this result is to be interpreted as modulo terms that will be of
higher order after renormalization and expansion in the RG invariant.

\section{Renormalization}

The results from the previous section are divergent, and we used
dimensional regularization to compute them. As advertised we make use of
the similarities between the free energy and the quantum effective action
to discuss the renormalization of the former, modeled by the
renormalization of the latter (see e.g. \cite{coleman}). In this section
we follow a more or less naive renormalization procedure, and compute the
renormalization group improved perturbative two-loop free energy. It will
be the role of the next section to tight the loose end, and prove that
what we did is indeed correct.

Since we are at second order of perturbation theory, we have both first
and second order poles in dimensional regularization, as can be seen in
eq. (\ref{fre2}). Since the free energy is a physical quantity, it has to
be well defined after renormalization. The recipe of this section is to
use the renormalization of the deformed principal \szm\
(\ref{couplren1}),(\ref{bf1}), and the renormalization of its dual, to
renormalize the free energy. The above renormalizations were performed
using the geometric method of \cite{alv,osborn}.

This procedure can be immediately objected since the free energy is
computed in a theory that has additional terms in the Lagrangian
(\ref{pertlagsub}), compared to the deformed principal \szm \
(\ref{lagrsig2}) and its dual (\ref{psilag}). A priori there is no reason
to expect that the wavefunction and coupling constant renormalization
functions are the same, though in fact they are. In the following we
present a simple argument in favor of the above statement. We shall give a
complete proof in section 5.5 below.

Coupling the external field amounts to the appearance of terms
proportional to $h$. Setting $h=0$ we get the original models (deformed
principal \szm \ and its dual). Thus the wavefunction and coupling
constant renormalization functions can at most differ from the ones of the
original models by terms proportional to $h$. But $h$ is a dimensionful
quantity, having mass dimension $+1$ (it is a super-renormalizable
coupling or relevant perturbation). On the other hand our theories have no
other dimensionful quantities, and the wavefunction and coupling constant
renormalization functions are dimensionless. We conclude that $h$ cannot
appear in the latter ones, proving the assertion.

In addition since $h$ couples to a conserved charge, it is not
renormalized. The only issue that remains will be to deal with the
renormalization of the operators that couple to $h$, viewed as composite
operators. We postpone this to section 5.5.

In order to cancel the second order poles of the regularized free energy
(\ref{fre2}) we have to go beyond the computation of \cite{bfhp}. Let us
review what we know at this stage about the renormalization of the
deformed principal \szm \ and its dual. In (\ref{lagrsigunif}) we have
introduced a generalized Lagrangian, and argued in the last paragraph of
the section that it encompasses the renormalization properties of both
models. Thus we can translate the renormalization properties of the
original models to those of the generalized Lagrangian. The Euler angles
($\phi$,$\theta$,$\psi$) are compact and have no wavefunction
renormalization. The renormalization of $r,\lambda$ and $a$ in the unified
model follows from those of $g$ (resp. $a$) and $\lambda$ in the original
models.

In (\ref{rescale}) we rescaled the fields ($\phi$,$\theta$,$\psi$), this
way in (\ref{lpert}) they have the corresponding nontrivial wavefunction
renormalization.  Thus $Z_\lambda(\lambda,M)$, the coupling constant
renormalization of $\lambda$, is the central object for their
renormalization. (Recall that $M$ is the renormalization group invariant
that appeared in (\ref{reninv})). $Z_\lambda(\lambda,M)$ was computed at
two loop order in perturbation theory in \cite{bfhp}. The interest in
\cite{bfhp} was restricted to the single pole terms. On the other hand we
are constrained to deal with the second order poles too. First we compute
the residue of the second order pole in $Z_\lambda(\lambda ,M)$.

\subsection{$\lambda$ at two loop} 

The goal is to determine the terms ${y}_\lambda(\lambda ,M)$ and
$\bar{y}_\lambda(\lambda ,M)$ in the
following expansion:
\be\label{cr}
Z_\lambda(\lambda ,M)=
1+{1\over \epsilon}\, y_\lambda(\lambda ,M)+{1\over
\epsilon^2}\, \bar{y}_\lambda(\lambda ,M)+{\tt o}({1\over
\epsilon^3}) .
\ee
The first term was basically determined in \cite{bfhp}. All we need is to
use (\ref{bf1}) for $Z_\lambda(\lambda ,g)$ and (\ref{reninva})
for $g$, in terms of the renormalization group invariant $M$, to obtain:
\be\label{cr1}
y_\lambda(\lambda ,M)=-{1\over {2\pi}}\lambda+{1\over {8\pi}}
(2\alpha_1^2-{1\over {\pi}})\lambda^2.
\ee
Our task of computing $\bar{y}_\lambda(\lambda ,M)$ is highly simplified
by the special properties of the non-linear \szm s. Following
\cite{thooft} it was shown in \cite{alv} that the generalized
renormalization theory \cite{Friedan} of non-linear \szm s lead to
generalized renormalization group equations, that allow one to determine
the residues of the higher order poles in a given coupling constant
renormalization function like $Z_\lambda(\lambda ,M)$, without extra 
diagrammatic computations.

More precisely it is shown that in a theory (like ours)  with a single
dimensionless coupling constant $\lambda$, with mass scale parameter
$\mu$, defined in $n$ dimensions (dimensionally regularized and minimally
subtracted), having an expansion of the bare coupling $\lambda_0$ in terms
of the renormalized coupling $\lambda$ of the form
\be\label{cr2}
\lambda_0=\mu^{2-n}\left(\lambda+\sum_{\nu=1}^\infty
\frac{a_\nu(\lambda)}{ 
(n-2)^\nu}\right),
\ee
the pole residues $a_\nu(\lambda)$ satisfy the recursive pole
equations:
\be\label{cr3}
(1-\lambda{\pa\over{\pa\lambda}})a_{\nu+1}(\lambda)= 
(1-\lambda{\pa\over{\pa\lambda}})a_1(\lambda){\pa\over{\pa\lambda}}
a_\nu(\lambda).
\ee

We can apply this directly to compute $\bar{y}_\lambda(\lambda ,M)$. From
the expressions following (\ref{couplren1}) we see that in our case
$a_1(\lambda)=\lambda\, y_\lambda(\lambda ,M)$, with $y_\lambda(\lambda
,M)$
given in (\ref{cr1}). Using (\ref{cr3}) for $\nu=1$ we can determine the
leading term in the expansion of $a_2(\lambda)$. A simple calculation
shows that $a_2(\lambda)={1\over {4\pi^2} } \lambda^3+\ldots$, implying
\be\label{cr4}
\bar{y}_\lambda(\lambda ,M)={1\over {4\pi^2} } \lambda^2+\ldots\,.
\ee

\subsection{One loop free energy}

By now it is a computation on the back of an envelope to obtain the
renormalized one loop free energy density. We have to use (\ref{frem2})
for the
leading term (with $\lambda$ as the bare coupling $\lambda_0$) and
(\ref{fre0}) for the one loop regularized result. The integral
itself, as it was pointed out in \cite{pall2}, is hard to
deal with exactly. Nevertheless, with the parameters
traded for the RG invariants, and keeping only the leading contribution in
$\lambda$ (as explained in the last paragraph of Section 4), and
renormalizing the expression using (\ref{couplren1}),
(\ref{cr}) and (\ref{cr1}), we obtain:
\be\label{f1}
{\cal F}^{1-loop}(h)=h^2\left( -{1\over{2\lambda}}-
\frac{1}{8\pi}\left[\ln\left(\frac{h}{\mu}\right)^{2}+\gamma
- 1 - \ln (4\pi)\right]\right).
\ee
Some of the terms that are higher order in $\lambda$, and are not yet
displayed, will be used for the computation of the ${\cal F}_2$.

We note two things here. Firstly, the one loop equivalence might be argued
to be not surprising based on the experience gained in \cite{en} and the
general one-loop beta function result of \cite{haagensen1}. Secondly, the
above expression correctly reproduces the known one-loop free energy of
the $O(3)$ sigma model, that arises in the $\alpha_1=0 $ limit.

As opposed to the corresponding computation in \cite{pall2}, the procedure
that leads to (\ref{f1}) does not require the splitting of the momentum
integration in the 0-th direction and the rest. The reason is simply that
due to the expansion in the RG invariant, the order $\lambda$ term
becomes totally covariant. The non-covariance is shifted to the next
order.

\subsection{Two loop free energy}

For the computation of the two loop free energy we need all the results
developed. The ${\tt o}(\lambda)$ corrections arising 
from the leading bare term
(\ref{frem2}) are obvious in the light of (\ref{couplren1}), (\ref{cr}), 
(\ref{cr1}) and (\ref{cr4}) . The ${\tt o}(\lambda)$ contributions 
from the bare next
to leading term of the one loop free energy (\ref{fre0}) can be computed
as
an expansion in the RG invariant, using (\ref{reninva})   
and (\ref{fundpar1}). The results are:
\be\label{l1}
\lambda h^2\frac{\alpha_1^{\,2}}{2}\,\left[-\frac{1}{4\pi\epsilon}
-\frac{1}{8\pi}(\,\gamma-\ln(4\pi)+2\,\ln (h)-1)\right]\,{\cal M}\,,
\ee
where ${\cal M}$ equals $\delta_{11}$ in the original model, and 
$\epsilon_{0\nu}\epsilon_{0\nu}$ in the dual one. Although, as argued
above, the product of two epsilon tensors with uncontracted indices is
ambiguous in dimensional regularization, the contraction of one index
gives a meaningful expression. Following \cite{epsilon}, we assume that
$\epsilon_{\mu\alpha}\epsilon_{\nu\alpha}$ has a consistent continuation,
namely: 
\be 
\epsilon_{\mu\nu}=-\epsilon_{\nu\mu}\,\quad
\epsilon_{\mu\alpha}\epsilon_{\nu\alpha}=\delta_{\mu\nu}.  
\ee

These two expressions might differ depending on the regularization schemes
chosen. As we have pointed out in connection with (\ref{dr}), we can a
priori continue the 0th direction into $n_0$ dimensions and the 1st
direction into $n_1$ dimensions, provided $n_0+n_1=n$. 
 We checked that the
final result is consistent with dimensional analysis in either case. While
we know \cite{pall2} that the choice $n_0=1$ is a consistent scheme with
the continuation of the $\epsilon$ tensor, we cannot claim the same about
the general case.

As we will see in a moment the only discrepancy in the two-loop free
energies of the two models comes from (\ref{l1}). This way we have several
choices of different schemes to see the quantum equivalence. With the
notation following (\ref{l1}) we can chose in both models the scheme with
$\delta_{00}=\delta_{11}$. The second choice is two different schemes in
the two models, but related to each other by:
$\delta_{00}^D=\delta_{11}^O$. Besides these naive choices we have a
highly non-trivial one, with the identical choice in both models:
$\delta_{00}=1$ and $\delta_{11}=n-1$. Unlike for the previously
mentioned ones, we know the consistency of this scheme, and we are going to
work with it in the rest of the paper. As we remarked in connection with
(\ref{fre0}), it was shown in \cite{pall2} that in this case the
difference in (\ref{l1}) is accountable for a perturbative redefinition of
the coupling constant $\lambda$ in the dual model. Naturally this
redefinition has to proceed the expansion in the RG invariant. Unlike in
\cite{pall2}, where this gave correction to ${\cal F}^{1-loop}(h)$, due to
the expansion in the RG invariant we get contribution to ${\cal
F}^{2-loop}(h)$. The redefinition of the coupling constant does not effect
the genuinely higher order terms as it can be seen from (\ref{psicoupl}).

Thus, as we just noted, we are going to use the last scheme in what
follows: $\delta_{00}=1$ and $\delta_{11}=n-1$, and for definiteness we
consider the case of the original model first. In the light of what we
just said, for the dual model, the results (\ref{f}), (\ref{ff}) below,
must be amended by $\frac{h^2}{8\pi}\lambda\frac{1}{\sqrt{M}}$, to account
for the perturbative redefinition of $\tilde{\lambda}$, eq.
(\ref{psicoupl}). However once this redefinition is taken into account,
the two loop results below are {\em identical} in the two models.  The
result for the other schemes will differ from the results to be presented
by terms that are some number times $-\lambda
h^2\frac{\alpha_1^{\,2}}{8\pi}$.

Using the bare result (\ref{fre2}) and
summing it with the corresponding contributions described, we obtain a
cancelation of both the first and second order poles, and obtain a finite
expression. The computation is somewhat tedious, but straightforward, the
result is simply:
\be\label{f}
\eqalign{
&{\cal F}^{2-loop}(h)=-{h^2\over{2\lambda}}-
\frac{h^2}{8\pi}\left[\ln\left(\frac{h}{\mu}\right)^{2}+\gamma - 1 - \ln
(4\pi)\right]-\cr
&-\frac{h^2\lambda}{16\pi^{2}}\left\{\left[\ln\left(\frac{h}{\mu}\right)^{2}
+ \gamma -\frac{1}{2} - \ln(4\pi)\right]
-\pi\alpha_1^{\,2}\left[\ln\left(\frac{h}{\mu}\right)^{2}+ \gamma- 1 - \ln
(4\pi)\right]\right\}.
}
\ee
It simplifies a bit if we use instead of minimal subtraction (MS
scheme) the
$\overline{MS}$ scheme: $\ln\mu\rightarrow\ln\mu+(\gamma-\ln (4\pi))/2$.
Then the free energy reads:
\be\label{ff}
\eqalign{
{\cal F}^{2-loop}(h)=&-\frac{h^2}{2\lambda}-\frac{h^2}{8\pi}\left[\ln
\left(\frac{h}{\mu}\right)^{2}-1\right]-\cr
&-\frac{h^2\lambda}{16\pi^{2}}\left\{\left[
\ln\left(\frac{h}{\mu}\right)^{2} - \frac
{1}{2}\right]-\pi\alpha_1^{\,2}\,
\left[\ln\left(\frac{h}{\mu}\right)^{2}-1\right] \right\}.
}
\ee

\subsection{Improvement of the perturbation theory}

We can take advantage of the asymptotic freedom of our theory and
calculate the RG improvement of the perturbative result.  This gives the
asymptotic expansion of the free energy for large values of the external
fields.

Physical quantities depend on the renormalized coupling $\lambda$,
the renormalized parameter $g$, and the dimensionful scale parameter
(or subtraction point) $\mu$, in such a way
that the action
of the renormalization group (RG) operator
\be
{\cal D}=\mu\,{\partial\over\partial\mu}
+\beta_\lambda(\lambda,g)\,{\partial\over\partial\lambda}
+\beta_g(\lambda,g)\,{\partial\over\partial g}\,,
\label{RGop}
\ee
vanishes on them. As the free energy density takes the form ${\cal
  F}(h)=-h^2\Psi (\lambda ,g,\mu ,h)$ and  
the external field $h$ is not
renormalized, the function $\Psi$ is renormalization invariant ${\cal
D}\Psi
=0$. Since we are interested in the behaviour of the free energy
density for large values of $h$ we write $h=h_0e^x$ where $h_0$ is
fixed and let $x\rightarrow \infty$. Standard RG considerations then
give that
\be\label{phrg}
{\cal F}(h)=-h^2\Psi (\lambda (x), g(x), \mu , h_0),
\ee 
where $\lambda (x)$ and $g(x)$ are the running coupling and parameter.
In our final result for the two loop free energy
density (\ref{ff}) the parameter $g$ is eliminated in favor of the
two loop RG
invariant.   As a consequence, up to this order, $\Psi =\Psi(\bar{\lambda} (x),\mu ,h_0)
$ depends only on $\bar{\lambda} (x)$,
which is a solution of  
\be 
{d\over dx}\bar{\lambda}(x)=\beta_\lambda(\bar{\lambda}(x),\alpha_1),
\quad \bar{\lambda}(x=0)=\lambda ,
\ee
where  the beta function has the following expansion:
\be
\beta_\lambda(\lambda)=-b_{0}\lambda^2-b_1\lambda^3-b_2\lambda^4+\ldots
\ee
with (see eq. (\ref{b1})) 
\be
b_{0}=\frac{1}{2\pi},\qquad
b_1=\frac{1}{4\pi}\left(\frac{1}{\pi}-\alpha_1^{\,2}\right)=\frac{1}{4\pi^2}\left(
  1-\frac{p}{2}\right),\quad p=\frac{2\pi}{\sqrt{M}}.
\ee 
Thus we can go on with the RG analysis as if we had only one coupling
constant $\bar{\lambda}(x)$. 
The expression of the (RG invariant) $\Lambda$ parameter is
\be
\frac{\Lambda}{\mu} =
e^{-\frac{1}{b_{0}\lambda}}\lambda^{-\frac{b_1}{b_{0}^{\,2}}
}e^{c}\left[1+\left(\frac{b_1^{\,2}}{b_{0}^{\,3}}
-\frac{b_2}{b_{0}^{\,2}}
\right)\lambda+{\tt o}(\lambda^2)\right],
\ee
or more conveniently
\be
\label{eff}
\ln\frac{\Lambda}{\mu}= - \frac{1}{b_{0}\lambda} -
\frac{b_1}{b_{0}^{\,2}}\ln\lambda
+c+\left(\frac{b_1^{\,2}}{b_{0}^{\,3}}
-\frac{b_2}{b_{0}^{\,2}}
\right)\lambda+ {\tt o}(\lambda^2),
\ee
where $c$ is a constant of integration. We define
$\Lambda\equiv\Lambda_{\overline{\rm MS}}$ by choosing 
$e^c=\left(\frac{1}{2\pi}\right)^{-\frac{b_1}{b_0^2}}$ as it
simplifies some of the forthcoming expressions.
 
As the next step of the RG analysis we note, that 
using expression (\ref{eff}), it is customary to define an {\em effective
coupling} $\alpha (h)$ by the following transcendental equation \cite{balog2}:
\be\label{alpdef}
b_{0}\,\ln\frac{h}{\Lambda_{\overline{\rm MS}}}
=\frac{1}{\alpha}+\frac{b_1}{b_0}\ln\alpha.
\ee
The important property of this effective coupling is that it depends
on the physical quantity $s=\ln\left(\frac{h}{\Lambda_{\overline{\rm MS}}}\right)$,
moreover one can express  the running coupling
$\bar{\lambda}$ in terms of the effective coupling $\alpha$ perturbatively:
\be\label{els}
\bar{\lambda} =\alpha\,(1+\xi_1\alpha+\xi_2\alpha^2+\ldots),
\ee
where
\be
\xi_1=b_0\left(\ln\frac{h_0}{\mu}-c\right),\qquad
\xi_2=\xi_1^{\,2}+\frac{b_1}{b_0}\xi_1-\frac{b_1^{\,2}}{b_{0}^{\,2}}
+\frac{b_2}{b_{0}}.
\ee
To use the ${\tt o}(\lambda )$ part of the free energy density
effectively we need the large $s$ asymptotic expansion of the
effective coupling up to ${\tt o}(s^{-3})$:
\be\label{uts}
\alpha =\frac{2\pi}{s}\left(1+\frac{A\ln s}{s}+\frac{B}{s}+
\frac{C\ln^2s}{s^2}+\frac{D\ln s}{s^2}+\frac{E}{s^2}\right)\,,
\ee
where 
the coefficients $A,\dots E$ can be obtained from
(\ref{alpdef}). 

Now improving (\ref{ff}) by the RG, i.e. using
eq. (\ref{els}-\ref{uts}) in (\ref{ff}, \ref{phrg}), results,   
when the dust settles,  the following formula for the 
asymptotic (large $s$) behaviour of the two loop
free energy:
\be\eqalign{
{\cal F}^{2-loop}(h)=-&{h^2\over 4\pi}\Bigl( s +(1-\frac{p}{2})\ln
    s-\frac{1}{2}+(1-\frac{p}{2})^2\frac{\ln s}{s}\cr &+
\frac{1}{s}\left[ (1-\frac{p}{2})^2-8\pi^3b_2+\frac{p-1}{4}\right]
+{\tt o}(\frac{\ln^2s}{s^2})\Bigr)\,.}
\ee
Unfortunately $b_2$ has not yet been computed in the literature,
nevertheless we can say a lot about it. Due to the expansion in the RG
invariant, it will have terms coming from the lower order beta function
coefficients, like in (\ref{bf1}), and the genuine three loop coefficient
evaluated at $g=-1$. For simplicity we denote this last term
$b^{(3)}=b^{({3-loop})}(g=-1)$. But this is exactly the three loop beta
function coefficients of the $O(3)$ $\sigma$-model: $b^{(3)}=b^{O(3)}_3$.
This beta function has been computed in \cite{brezin}, and in our
convention for the coupling constant $b^{(3)}$ takes the form
$b^{(3)}=\frac{5}{32\pi^3}$.  
Carrying one step further the computation of
(\ref{b1}) results:
\be
b_2=\frac{1}{8\pi}\left(\alpha_1^{\,4}-\frac{3}{\pi}\alpha_1^{\,2}\right)
+b^{(3)}=\frac{1}{16\pi^3}\left(\frac{p^2}{2}-3p\right)+\frac{5}{32\pi^3}
\,.
\ee
Accordingly, as a final result we have:
\be
{\cal F}^{2-loop}(h)=-{h^2\over 4\pi}\Bigl( s +(1-\frac{p}{2})\ln
    s-\frac{1}{2}+(1-\frac{p}{2})^2\frac{\ln s}{s}+
\frac{1}{s}\frac{3p-2}{4}
+{\tt o}(\frac{\ln^2s}{s^2})\Bigr)\,.
\ee

\subsection{Renormalization of the composite operators}

In this section we give a solid foundation to the results of the previous
renormalization procedure. More precisely we consider the renormalization
of the composite operators that arise via the coupling of the chemical
potential. Our attitude is similar to the standard procedure of mass
renormalization in QCD: initially the mass of the light quarks is set to
zero, then the bilinear mass operator $\bar{\psi}\psi$ is added, and its
effect is accounted by its renormalization as a composite operator. We
followed the same ideology so far, neglecting the fact that the
renormalization of the coupled theory is different from that of the
uncoupled. It is now that we remedy this.

Let us reconsider the computation of the free energy (\ref{freen}). In
(\ref{freenex})  we started with the bare Lagrangian. Instead we have to
use the renormalized one, with counterterms coming from the wavefunction
renormalization of the fields, coupling constant renormalization, and the
renormalization of the composite operators, all these coming with a
natural
grading:
\be\label{rf1}
S[h,\varphi]+\Delta S[h,\varphi]= \sum_{i=-2}^\infty
\left(S_i[h,\varphi]+\Delta S_i[h,\varphi]\right)\,A^i\,,
\ee
with the renormalized quantities in the right hand side. 
$\Delta S_i[h,\varphi]$ commonly denotes all the counterterms of order
$i$.
The original
Lagrangian that has to be employed reads:
\be\label{rf11}
{\cal L}^0=\frac{1}{2}(\pa_\mu\psi)^2+ \frac{1}{2}(\pa_\mu\phi)^2+ 
\frac{1}{2}(\pa_\mu\theta)^2
+A\,a\omega_{\mu\nu}\,\theta\pa_\mu\phi\pa_\nu\psi
+A^2\frac{1}{2}r\theta^2(\pa_\mu\phi)^2\,.
\ee
$\Delta S_{-2}$ amounts simply to the multiplicative renormalization we
considered in (\ref{cr}). Since multiplied by a factor of 
$1/\lambda$, it gives
rise to a term that is independent of $\lambda$ (already used to renormalize 
the one loop free energy), one that is
${\tt o}(\lambda)$ (also used), and higher order terms:
\be\label{rf2}
\Delta S_{-2}=\Delta S_{-2}^0+\Delta
S_{-2}^1\,\lambda+{\tt o}(\lambda^2).
\ee
$\Delta S_{-1}=0$ as already $S_{-1}=0$. A priori $\Delta S_0$ contains
terms from the from the wavefunction renormalization of the fields, but
these are independent of $h$, and are canceled by the denominator in
(\ref{freen}). Thus $\Delta S_0=\delta_{{\cal O}^0}\,{\cal
O}^0+\delta_{{\cal O}^1}\,{\cal O}^1$, where ${\cal
O}^0=-r\frac{1}{2}h^2\theta^2$ and ${\cal
O}^1=ah{\Omega}_{\nu}\,\theta\pa_\nu\psi$. The renormalization of ${\cal
O}^0$ (and the similar operators that appear in the Lagrangian
proportionally to $h$) has two contributions: one from the renormalization
of $h$ (that is zero as discussed in the previous section) and one from
the renormalization of the composite operator $\theta^2$. We expect
that $\delta_{{\cal O}^0}=1+{\tt o}(\lambda^2)$, and this expectation
can be confirmed by a short explicit computation.

In order to compute $\delta_{{\cal O}^0}$ we need a Green's function
involving ${\cal O}^0$. For simplicity let's consider the one-point
function of $\theta^2$: $\langle \theta^2(x)\rangle $. As a
renormalization prescription we normalize the one-point function according
to the tree level value, and determine $\delta_{\theta^2}$ from the
condition of preserving the above normalization. At ${\tt o}(\lambda)$
there are two diagrams that contribute coming from the ${\tt o}(A)$ and
${\tt o}(A^2)$ vertices. The latter one has a value proportional to

\be
\int\! d^ny\,G(x-y)^2\,\pa^2G(0)=\int\! d^ny\,G(x-y)^2\,\delta(0)=0
\ee
in dimensional regularization. The same result is obtained if the diagram
is evaluated in momentum space, where the masslessness of the fields
requires additional IR regularization (as exploited already in a
previous section). 

The other diagram is readily proportional to $\lambda\,a^2$. The remaining
integral can be computed but we don't need the result for what follows,
because once again, as we express this contribution in terms of the RG
invariants we shall have a dependence proportional to $\lambda^2$. Thus we
have $\delta_{\cal O}=1+{\tt o}(A^3)$ at least. As a consistency check we
quote that the same result is obtained by considering for example the
$\langle \theta^2(x)\theta(y)\theta(y')\rangle$ Green's function.

During the computation we employed the $\frac{1}{2}h^2\theta^2$ operator
as a mass term for $\theta$, though it is a composite operator as
discussed above. The motivation for this can be given as follows: assume
that $\frac{1}{2}h^2\theta^2$ is a perturbation, and expand it
perturbatively with the rest of the terms in (\ref{inerm}). The difference
is that $\frac{1}{2}h^2\theta^2$ is independent of $\lambda$, and as such
any term in its power series expansion is of the same order, and has to be
summed. In other words, any term in the perturbative expansion of ${\cal
F}$ will be multiplied by the full expansion of the exponential of
$\frac{1}{2}h^2\theta^2$, that can be resummed. The resummation on the
other hand is equivalent with the corresponding mass term $h$ for the
$\theta$ field. From the point of view of the original $h=0$ theory this
is a non-perturbative result.

Next we consider the renormalization of the second operator ${\cal O}^1$
and the one appearing at the next level ${\tt o}(A)$:  ${\cal
O}^2=\theta^2\pa_0\phi$. These operators will have their composite
operator renormalization functions: $\delta_{{\cal O}^1}$ and
$\delta_{{\cal O}^2}$. Since the undeformed ($h=0$) theory has only
interactions proportional to $\lambda$ or to $a\sqrt{\lambda}$, and the
latter ones must appear at least twice for a finite contribution, eq.
(\ref{rf11}), we conclude that the renormalization effects due to these
vertices are at least of order $\lambda$ and resp. $\lambda\,a^2$:
$\delta_{{\cal O}^1}=1+z^1\,a^2\lambda+\ldots$ resp. $\delta_{{\cal
O}^2}=1+z^2\lambda+\ldots$.  In the spirit outlined above we have to
introduce the new terms $z^1{\cal O}^1$ and $z^2{\cal O}^2$ into the
action (\ref{rf1}) as terms contributing to $\Delta S[h,\varphi]$, and
account for their contribution. As these new terms are proportional to
$\lambda$, they contribute to $\Delta S_2[h,\varphi]$. Thus we expect new
contribution to ${\cal F}_2$. Based on (\ref{inerm}) these are
proportional to $\langle z^1{\cal O}^1\rangle$ resp. $\langle z^2{\cal
O}^2\rangle$. But we already know from (\ref{fre1}) that these are zero.
It is easy to see that all the other combinations of these two operators
with the rest of the operators will give higher order contributions to
the free energy.

It is obvious that the product of $z^1{\cal O}^1$ (resp. $z^2{\cal O}^2$)
with ${\cal O}^1$ (resp. ${\cal O}^2$) will give non-zero contribution to
${\cal F}_3$, but as far as ${\cal F}_2$ is concerned there is no
deviation from the results we obtained.

At the next level (i.e. at ${\tt o}(A^2)$) we have the composite operators ${\cal
O}^3=\theta^4$ and ${\cal O}^4=\theta^3\pa_\nu\psi$. Consider first ${\cal
O}^3=\theta^4$. In order to compute $\delta_{{\cal O}^3}$ we fix the
normalization of the correlation function $\langle \theta^4(x)\rangle$ to
the tree level value: $\langle \theta^4(x)\rangle=3\bar{G}(0)^2$. We
compute $\delta_{{\cal O}^3}$ by computing the first corrections to the
correlation function, and insisting as above that the above normalization
remains valid. Using the available interaction terms in (\ref{rf11}) we
see that using the vertices $\theta\pa_\mu\phi\pa_\nu\psi$ resp.
$\theta^2(\pa_\mu\phi)^2$ give the first nontrivial contribution
proportional to $\lambda^4$ resp. $\lambda^2$, while their combination is
of order $\lambda^3$. Thus we conclude that $\delta_{{\cal
O}^3}=1+{\tt o}(\lambda^2)$. In the case of ${\cal O}^4$ after simple
considerations it is similarly obtained that $\delta_{{\cal
O}^4}=1+{\tt o}(\lambda^2)$, and we are done.

From the above analysis we conclude that the more or less naive 
renormalization procedure employed in the first place is completely
adequate, and the additional composite operator renormalizations do not
disturb the results as far as the ${\cal F}_2$ is concerned.

\section{Conclusions}

In this paper we have computed the change in the ground state energy
density of a deformed principal $SU(2)$ sigma model, and one of its
T-duals, in the presence of an external field. The computations has been
carried out at two loop order in perturbation theory. Perfect agreement
has been found in the following sense: there were renormalization schemes
in the two models that yielded the same expressions for the two loop free
energy densities.

We defined the free energy using the conserved charge corresponding to a
special symmetry of the model. Care had to be made at the choice of the
charge, since we wanted it to exist as a genuine symmetry in both the
original and dual model, while it is known that the charge corresponding
to the isometry used for the duality becomes topological. We performed the
computation for the simplest choice, but other possibilities are also
investigated at least in the one loop order \cite{pall2}, \cite{balog}.  
We computed the relevant diagrams and performed the renormalization using
dimensional regularization.  Due to asymptotic freedom, we could improve
the perturbative result using the renormalization group equations.

This result strengthens the confidence in the findings of \cite{bfhp},
that were mainly based on beta function computations.  More precisely this
way we have given much stronger evidence that in the case of constant
$g_{00}$, at least in the case of the deformed principal $SU(2)$ sigma
model, the naive Buscher formula that relates the original model and its
T-dual, gives a true quantum equivalence. As it was pointed out and
exemplified in \cite{en}, the constancy of $g_{00}$ is no guarantee for
two loop quantum equivalence. Based on this one might expect that a finer
test would detect discrepancy even in the case where the beta function
arguments show no sign of it.  Though the free energy test that we
performed does not prove the lack of discrepancies, it hints to the
non-existence of these at least at two loop order. It would be interesting
to analyze the pertinent $SL(3)$ example of \cite{en} from the free energy
point of view, though, at this point it seems to be too big of a
computational challenge.

\bfl
{\large \bf Acknowledgments}
\efl

We would like to thank J. Balog, P. Forg\'acs for valuable discussions
throughout this work. R.L.K. would also like to thank F.P. Esposito, D.
Kastler, D. Maison, F. Mansouri, T. Nagy, M.E. Peskin, G. P\'ocsik, A
Rebhan, A. Slavnov, P. Sur\'anyi, L.C.R. Wijewardhana, and L. Witten for
discussions at different stages of this work. This work was supported in
part by the Hungarian National Science Fund (OTKA) under T029802, and by
the Ministry of Education under FKFP 0178/1999. R.L.K. was also supported
in part by DOE grant DOE-FGO2-84ER40153.

\ed